\documentclass[12pt]{iopart}
\usepackage{graphicx}
\usepackage{epstopdf}
\usepackage{multirow}
\usepackage{subfig}
\usepackage{cite}
\usepackage{threeparttable}
\usepackage{float}

\begin{document}

\title[]{A practical applicable quantum-classical hybrid ant colony algorithm for the NISQ era}

\author{Qian Qiu$^2$$^,$$^4$, Liang Zhang$^2$$^,$$^4$, Mohan Wu$^1$, Qichun Sun$^2$, Xiaogang Li$^2$, Da-Chuang Li$^{3\dagger}$, Hua Xu$^{1\dagger}$}
\address{$^1$College of Artificial Intelligence, Tianjin University of Science and Technology, Tianjin 300457, China}
\address{$^2$ Yiwei Quantum Technology Co., Ltd, Hefei 230088, China}
\address{$^3$School of Physics and Materials Engineering, Hefei Normal University, Hefei, Anhui 230601, China}
\address{$^4$ These authors contributed equally: Qian Qiu, Liang Zhang}
\ead{hua.xu@ywquantum.com (X. Hua), dachuangli@ustc.edu.cn (L. Da-Chuang)}
\vspace{10pt}
\begin{indented}
\item[]September 2024
\end{indented}

\begin{abstract}
Quantum ant colony optimization (QACO) has drew much attention since it combines the advantages of quantum computing and ant colony
optimization (ACO) algorithm overcoming some limitations of the traditional ACO algorithm.
However, due to the hardware resource limitations of currently available quantum computers, the practical application of the QACO is
still not realized. In this paper, we developed a quantum-classical hybrid algorithm by combining the clustering algorithm with QACO
algorithm.This extended QACO can handle large-scale optimization problems with currently available quantum computing resource. We
have tested the effectiveness and performance of the extended QACO algorithm with the Travelling Salesman Problem (TSP) as
benchmarks, and found the algorithm achieves better performance under multiple diverse datasets. In addition, we investigated the noise impact
on the extended QACO and evaluated its operation possibility on current available noisy intermediate scale quantum(NISQ) devices.
Our work shows that the combination of the clustering algorithm with QACO effectively improved its problem solving scale, which
makes its practical application possible in current NISQ era of quantum computing.
\end{abstract}

%
\vspace{2pc}

\section{Introduction}
With the exponential growth of data and complexity of problems, efficiently solving large-scale combinatorial optimization problems has become a significant research focus. The large scale combinatorial optimization problem is a subclass of the combinatorial optimization problem, involving the search for an optimal solution within a large and discrete solution space. The main characteristic of the combinatorial optimization problems is that the solution space is discrete rather than continuous, which makes traditional continuous optimization methods (such as gradient descent algorithm) inapplicable. One of mainstream methods for solving large-scale combinatorial optimization problems is
heuristic algorithms, which can find solutions close to the global optimum within a relatively short runtime.
The Ant Colony Optimization (ACO) algorithm, one of the most representative heuristic algorithms, inspired by the behavior of ants
foraging\cite{Dorigo,Ghosh}, can rapidly find
near-optimal solutions in complex search spaces and has been widely applied to the Traveling Salesman Problem
(TSP)\cite{Maity,BLi,Mohsin}, the Quadratic Assignment Problem\cite{Demiral,Amirhosein,Alfredo}
and the Vehicle Routing Problem (VRP)\cite{Madhloom,YDing,LAn,TLi}, etc. Despite its many virtues, the ACO algorithm still
confronts challenges in tackling larger combinatorial optimization problems, such as slow convergence and high computational
resource requirements.

Quantum computing, as an emerging computational paradigm, leverages quantum superposition and entanglement to provide new
opportunities for solving optimization problems\cite{Benioff,Deutsch,Shor}. With significant progress achieved over the past
decades, quantum computing is currently in the Noisy Intermediate-Scale Quantum (NISQ) era, transitioning from scientific research to
industrial applications\cite{ZewenLi,Wojciech,BAlhijawi,Musatafa,SKMaruf,XLMa}. Over the past decades, a few quantum-classical hybrid
algorithms have been proposed to improve the efficiency of the ACO algorithm, such as using quantum circuits to generate true random
numbers to prevent getting trapped in local optima, as well as introducing novel coding strategies through quantum gates.
Ling Wang {\it et al.} first proposed a novel pheromone updating strategy that utilized quantum rotation gates to refresh the
pheromone
depending on the iterative result in 2007\cite{QNiu}. In their work, they focused on addressing the issue of iteration results
stagnating around local optima. In 2010, Panchi Li {\it et al}. proposed a
QACO algorithm that codified the position of each ant with many qubits\cite{Panchi}. The algorithm collected the
local best solution depending on pheromone trail on each path by utilizing quantum gate to update every ants representative qubits. To
induce path mutation and extend the capacity of the solution pool, this algorithm incorporated quantum non-gates operations. In 2022, M. Garcia de Andoin and J. Echanobe proposed an hybrid quantum ACO algorithm that was using quantum computing for tackling pheromone updating and exploration parameter, then passing the processing result to a classical computer for the rest parties of this algorithm\cite{Echanobe}. However, due to the limitations in both the quantity and quality of available
qubits in real quantum computers, the usefulness and practical application of the algorithm have not been tested and verified on
large scale problems.

In this work, we introduces a quantum-classical hybrid ant colony optimization algorithm (QACO) combined with the K-means clustering method. The core idea of this approach is to divide large-scale TSP problems into multiple smaller subgroups using K-means
clustering, then employ quantum computers to find optimal solutions for these subgroups, and finally refine the global
solution through parallel ACO. By adopting this approach, the extended QACO effectively addresses the computational challenges of
large-scale TSP problems under the current limitations of quantum computing resources, expanding the practical application scenarios
for QACO.

The main contributions of this paper are as follows:\\
1. A quantum-classical hybrid QACO algorithm that combines K-means clustering is proposed, addressing some of the hardware
limitations of current quantum computers.\\
2. The algorithm has been tested on multiple datasets, demonstrating its effectiveness and superior performance in solving
combinatorial optimization problems.\\
3. The algorithm's noise resistance is assessed during the computation process, demonstrating its potential for application in the
NISQ era.

The rest of this paper is organized as follows: Section 2 introduces the basic principles of the classical ACO, quantum circuits
and the K-means clustering method. Section 3 details the steps of the extended QACO algorithm.
Section 4 presents our experimental results and analysis. Finally, the last section concludes the paper and discusses future
research directions.

\section{ACO and K-means Clustering algorithms}
\subsection{Ant Colony Optimization Algorithm}
The ACO algorithm is a meta-heuristic inspired by the foraging behavior of real ants. Its fundamental principle is to emulate the
ability of ant colonies to discover the shortest paths between their nests and food sources. As individual ants explore their
surroundings, they leave behind pheromone trails. These trails serve as communication channels, directing other ants toward the most
promising path.
The ACO algorithm mimics this natural behavior by employing a population of artificial ants to search for optimal solutions within
a given problem space.

Initially, ants are randomly distributed on each node and move to any reachable node. During the movement process, ants generate
pheromone trails. To expand the solution space and prevent stagnation in local optima, the ACO algorithm introduces an evaporation
mechanism. After each iteration, the pheromone is updated to influence the probability of each path being chosen by ants in the next
iteration.

The previously mentioned pheromone updating process can be summarized as the ant decision rule, which uses a series of guidelines to
influence an ant's stochastic choice of the next feasible node. At time t, the k-th ant located on the node r moves to the next
node s based on the following rule.

\begin{equation}
s = \left \{
\begin{array}{ll}
  \mathop{argmax}\limits_{ru\in allowed_k(t)} \left [ \tau_{ru}(t)^{\alpha} \eta_{ru}^{\beta } \right ] &\mbox{when}\ q\le q_0 \\
  S &\mbox{otherwise}
\end{array}
\right .
\end{equation}

Where, $\tau_{ru}(t)$ represents the pheromone value at time $t$, $\eta_{ru}$ denotes the specific problem requirement, $\alpha$
is the impact factor, $q$ indicates a randomly number in $[0,1]$, $q_0$ is a constant number selected in $[0,1]$,
$allowed_k(t)$ denotes the eligible nodes set for the k-th ant at time t, $S$ is a parameter selected from $allowed_k(t)$
based on the following equation.

\begin{equation}
P_{rs}^k(t)= \left \{
\begin{array}{ll}
\frac{\tau_{rs(t)}^{\alpha}\eta_{rs}^{\beta}}{\sum\limits_{ru\in allowed_k(t)}\tau_{rs}(t)^{\alpha}\eta_{ru}^{\beta}} &\mbox{if}\ s \in allowed_k(t)\\
0 &\mbox{othewise}
\end{array}
\right.
\end{equation}

\subsection{K-means Clustering}
The K-means clustering algorithm is a widely used unsupervised machine learning technique for partitioning data into distinct groups
or clusters. The principle behind the K-means algorithm is to group data points based on their similarity, with the goal of minimizing
the sum of squared distances between data points and their respective cluster centroids\cite{Ikotun}. The algorithm begins by randomly
initializing K cluster centroids in the feature space, where K represents the desired number of clusters. Each data point is then
assigned to the nearest centroid based on a distance metric, typically Euclidean distance. The initial assignment creates the first
clustering of the data.

Next, an iterative process starts to refine the clustering. The algorithm updates the centroids by computing the mean of all data
points assigned to each cluster. This step recalculates the centroid positions, effectively shifting them towards the center of their
respective clusters. Subsequently, the data points are reassigned to the closest centroids based on the updated positions. These
centroid updates and reassignments continue iteratively until convergence, where the centroids no longer change significantly or a
predetermined number of iterations is reached.

K-means clustering aims to minimize the within-cluster sum of squares\cite{Konstantinos}, also known as the inertia or distortion. The
inertia represents the sum of squared distances between each data point and its assigned centroid. By minimizing the inertia, the
algorithm ensures that data points within each cluster are similar to each other, while distinct from data points in other clusters.
The performance and effectiveness of the K-means algorithm heavily depend on the initial centroid positions. Different initializations
can lead to different clustering results. To mitigate this issue, the algorithm is often run multiple times with different random
initializations, and the clustering result with the lowest inertia is selected as the final solution.

K-means clustering has various applications in data analysis, pattern recognition, image segmentation, customer segmentation, and
more. It provides a straightforward approach to uncovering underlying structures or patterns in data without requiring prior knowledge
or labeled training samples. Despite its popularity, the K-means algorithm has certain limitations. One major challenge is determining
the optimal number of clusters, K, which is often subjective and problem-dependent. Choosing an inappropriate value for K can lead to
sub-optimal or misleading clustering results. Additionally, the algorithm is sensitive to outliers and can be influenced by the
initial centroid positions, making it susceptible to converging to sub-optimal solutions.

To enhance the performance of K-means clustering, various extensions and variations have been proposed. These include initialization
techniques like K-means++ to improve the initial centroid selection\cite{SVassilvitskii}, and algorithms like K-means with mini-batch
updates
to handle large-scale data sets efficiently. Additionally, efforts have been made to address the limitations of K-means, such as
incorporating robust distance metrics, handling categorical data, as well as incorporating constraints into the clustering process.
While it has its limitations, advancements and adaptations have been made to address these challenges, making K-means a valuable tool
for data exploration, segmentation, and pattern recognition tasks.

\section{Quantum Ant Colony Optimization with K-means clustering method}
The application scope of quantum computing algorithm, however, is still confined due to the factors such as the scale of qubits and their fidelity. These limitations pose significant challenges to the widespread adoption of quantum computing algorithms.

Generally, if the computational resources are limited, a common approach to resolve the large scale problem is to partition it into a
smaller one that could be solved with available resources. In this article, the K-means clustering method is utilized to decompose a
large scale issue into a smaller one. In our proposed extended QACO, the qubits are categorized into two classes, the path generating qubits and ancilla qubit.

\begin{figure}[htbp]
\centering
\includegraphics[scale=0.5]{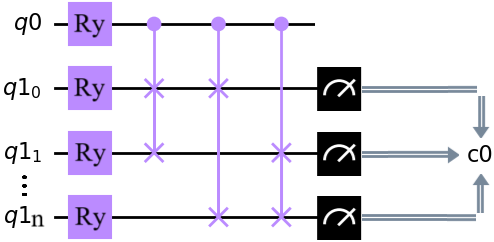}
\caption{The path search quantum circuit. In this diagram, $q0$ indicates the auxiliary qubit (ancilla) which make the solution to
mutate in the certain condition controlled by Ry gate and c0 denotes classical bits. Meanwhile, $q1_{0}$-$q1_{n}$ denote n+1
qubits for points in the combinatorial optimization problems.}\label{QCircuit}
\end{figure}

In the extended QACO algorithm, the quantum circuit diagram is shown Fig. \ref{QCircuit}. In Fig. \ref{QCircuit}, $q0$ denotes the
auxiliary qubit (ancilla) which make the generated path to mutate when the global best path remains invariant across many iterations.
The Ry gate is used to control the mutation probability. The $q1_i$ are used as a path generator that could codify the points in the
combinatorial optimization problems.
In the extended QACO algorithm, we exploit 9 qubits to address various
combinatorial optimization problems. One qubit is used as an auxiliary bit, while the other eight are used to encode the cities in the
TSP problem, with each two bits representing one city.
Since the current global solution during the process is not related to any known pattern, we introduce a random rotation angle within
$[0, \frac{\pi}{2}]$ as the mutation probability.

As is known, quantum computing measurements are statistical. Since the outcome of a quantum circuit may produce invalid results that
do not adhere to TSP rules, the extended QACO must include a result verification module. If the current solution during the iteration
is
infeasible, the algorithm generates a feasible path randomly during the first 10 iterations. Subsequently, the Hamming distance is
incorporated into the algorithm. The probability of an infeasible solution is set to be inversely proportional to the Hamming distance
from the existing solution pool. This probability is defined as follows:

\begin{equation}
p_i = \left(d_i\sum \limits_{j} \frac{1}{d_j}\right)^{-1}
\end{equation}

where, $d_i$ denotes the Hamming distance between the current infeasible solution and one of the best path in the solution pool, $j$
is the number of the solution pool.

Even though the extended QACO algorithm incorporates quantum computing, it still requires termination criteria to exit the search
loop.
In the proposed algorithm, we introduce two independent stopping conditions. One is to limit the number of iterations, and the other
is to set a convergence criterion where the search stops if no better results are found within a specified number of iterations.
To compare the performance differences between the proposed algorithm and the classical ACO algorithm, the classical ACO was applied
to the benchmarks with 6 ants, pheromone and distance factors set to 4 and 2 respectively, and performed 1000 iterations.

The main steps in the QACO with K-means clustering method can be described as below:\\
{\bf Step 1:} Utilizing K-means clustering to divide the larger problem into a smaller one that could be processed by the current
quantum computing resource, where the group numbers of the K-means clustering is relative to the applicable number of qubits. Here,
it is 4.\\
{\bf Step 2:} Initializing the parameters of the QACO, setting the number of generations and the rotation angel of the pheromone to
$\pi/2$, which means all solutions are generated with the same probabilities at the beginning of the algorithm.\\
{\bf Step 3:} Calculating the best solution in the iteration and comparing it with the global best solution. If the current best
solution is better than the global best solution, then substitute with the current best solution.\\
{\bf Step 4:} Determining whether the current iteration meet the termination condition. If not, going to the next step.\\
{\bf Step 5:} Updating the pheromone value with the quantum rotation gate.\\
{\bf Step 6:} Determining whether the scale of the clustering group is suitable for the number of the qubit. If not, continuing to
clustering and repeating the foregoing steps until the problem has be solved.

\begin{figure}[htbp]
\centering
\includegraphics[scale=0.5]{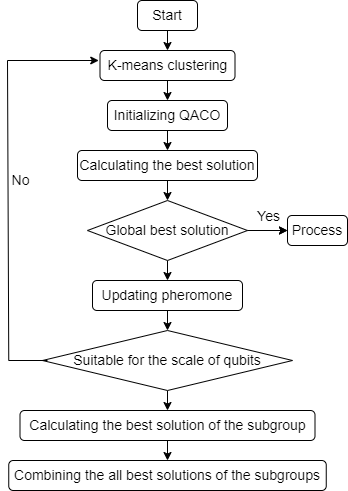}
\caption{The work flow of the extended QACO algorithm}\label{workflow}
\end{figure}

Figure \ref{workflow} shows the concise work flow of the proposed algorithm.

As the ACO, the updating pheromone strategy of QACO is also an important step. The updating pheromone table is given in table \ref{tab2}.

\begin{table}[h]\centering
\caption{The lookup table of the rotation angle of QACO}\label{tab2}
\begin{tabular}{cccc}
\hline
$x_i$ & $b_i$ & $f(x) > f(b)$ & $\Delta\theta_i$                   \\ \hline
0 & 0 & True                  & $-0.01\pi^{\ast}$ \\
0 & 0 & False                 & $0.04\pi $ \\
0 & 1 & True                  & $-0.05\pi^{\ast}$ \\
0 & 1 & False                 & $0.07\pi $ \\
1 & 0 & True                  & $0.05\pi^{\ast} $ \\
1 & 0 & False                 & $-0.07\pi$ \\
1 & 1 & True                  & $0.01\pi^{\ast} $ \\
1 & 1 & False                 & $-0.04\pi$ \\ \hline
\end{tabular}
\end{table}
\normalsize

Where $x_i$ means the state of the i-th qubit on the current iteration, $b_i$ indicates the global best state of the i-th qubit so far. $f(x)$ and $f(b)$ demonstrates the value of the explicit fitness function based on the specific problem, the asterisk denotes the
cosine value of $\Delta\theta_i$ less than $-1$ then it must be multiplied with $-1$.
\section{Experimental Results}
To evaluate the performance and practicality of the QACO algorithm with K-means clustering, we applied it to several datasets,
including ulysses-16, ulysses-22, eil-51, eil-76, berlin-52, bayg-29, as well as randomly generated data\cite{ReineltG}. These
datasets were selected as they are commonly used in combinatorial optimization, particularly for the Traveling Salesman Problem (TSP).
Ulysses-16 and Ulysses-22 are small-scale datasets that offer an initial test bed for algorithm validation. Eil-51, Eil-76, and
Berlin-52 are medium-scale datasets derived from real-world European city routes, providing a more complex environment for performance
testing. Bayg-29, based on 29 German cities, represents another small-scale, real-world problem. The inclusion of randomly generated
data ensures that the algorithm's robustness and adaptability can be assessed beyond these structured datasets. These tests were
conducted to verify the algorithm's key features and effectiveness. The results of the tests are given below. Fig. \ref{ULYresults}.
provided the result of Ulysses-16. All the numerical tests are performed by Yiwei Quantum Computing Simulation Platform.

The results of benchmarks we studied are summarized in table \ref{tab3}. In table \ref{tab3}, the first column shows the optimum of
all benchmarks while second and third column presents the results calculated by the classical ACO and the extended QACO, respectively.
The result of ulysses-16 is 62.14, a 20.17\% improvement of the solution given in recent published article\cite{MeijiaoL}.
The best distance calculated by classical ACO is 69.67 that is longer than that calculated by the extended QACO.
The results of ulysses-16 and bayg-29 calculated by the extended QACO and ACO are shown in figure \ref{ULYresults}.
To investigate whether the improvement is influenced
by the K-means clustering method, we introduce the clustering algorithm into the classical ACO. The result is 66.34
that is also longer than that of the extended QACO. It is indicated that the advancement of the proposed QACO is not induced by the
clustering method.

Besides Ulysses-16, the rest results of benchmarks in table \ref{tab3} are all longer than the optimal values.
However, the
results calculated by the extended QACO algorithm are significantly better than those of the classical ACO algorithm, indicating that
the extended QACO algorithm outperforms the classical ACO algorithm. The reason why the results calculated by the extended QACO
algorithm is longer than the optimal value is most likely due to the limited number of qubits, which restricts the performance of the
K-means clustering algorithm, i.e. the best clustering results for the data sets are not 4 that is determined
by the quantum circuits shown in the figure \ref{QCircuit} which could only process problems of size less than 4.


\begin{figure}[htbp]
\centering
\subfloat
{
\includegraphics[scale=0.5]{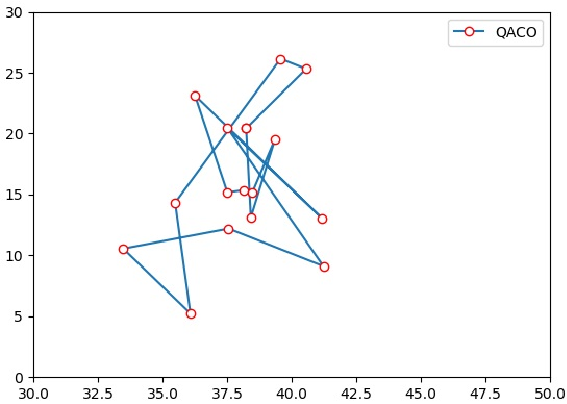}
}
\subfloat
{
\includegraphics[scale=0.5]{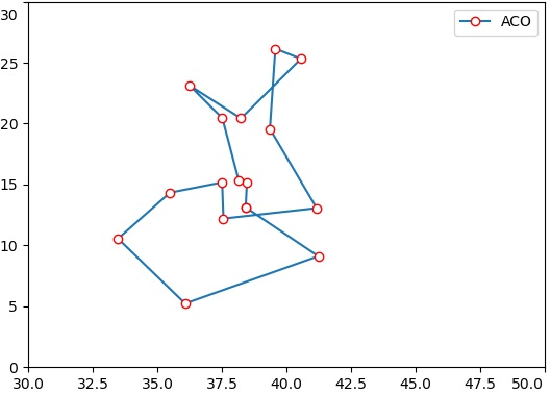}
}\\
\subfloat
{
\includegraphics[scale=0.475]{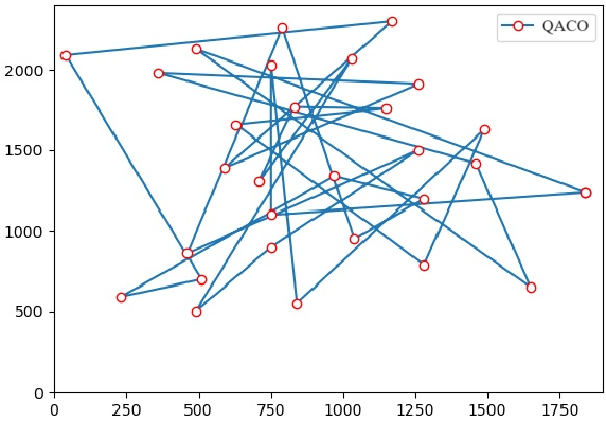}
}
\subfloat
{
\includegraphics[scale=0.35]{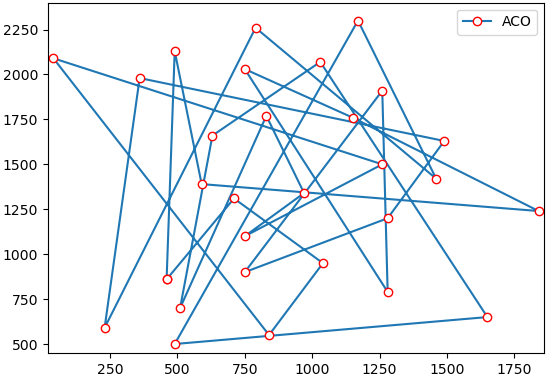}
}
\caption{The best path of the algorithm for solving ulysses-16 and bayg-29: First line:(left) QACO for ulysses-16 (right) ACO for
ulysses-16; Second line: (left) QACO for bayg-29 (right) ACO for bayg-29}\label{ULYresults}
\end{figure}

\begin{table}[h]\centering
\caption{The total distances of the numerical tests by the ACO and QACO}\label{tab3}
\begin{threeparttable}
\begin{tabular}{cccc}
\hline
                       & Optimum          & ACO       & QACO       \\ \hline
ulysses-16             & 77.84$^\dagger$  & 69.67     & 62.14     \\
ulysses-22             & 70.13$^*$        & 99.87     & 92.49     \\
bayg-29                & 9073$^\dagger$   & 14624.15  & 12545.95  \\
eil-51                 & 426$^*$          & 587.31    & 541.92    \\
eil-76                 & 538$^*$          & 1039.15   & 713.39    \\
berlin-52              & 7542$^*$         & 13450.71  & 9158.34   \\
random data(64 points) &   -              & 331187.31 & 148170.56 \\ \hline
\end{tabular}
\begin{tablenotes}
\footnotesize
\item[*] The optimum data are from TSPLIB
\item[$\dagger$]The optimum of ulysses-16 and bayg-29 from reference \cite{MeijiaoL} and \cite{HJinnai}
\end{tablenotes}
\end{threeparttable}
\end{table}

\begin{figure}[htbp]
\centering
\subfloat
{
\includegraphics[scale=0.305]{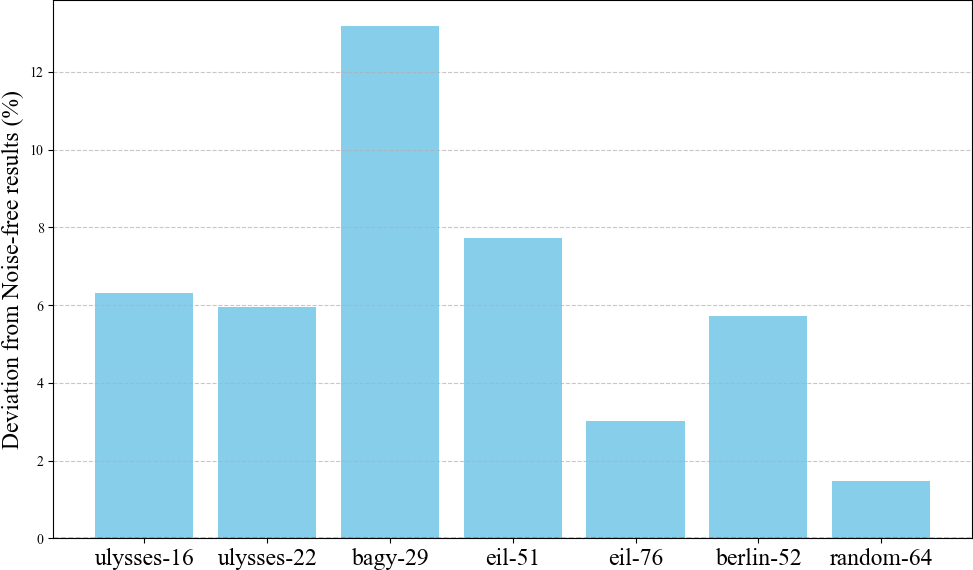}
}
\subfloat
{
\includegraphics[scale=0.305]{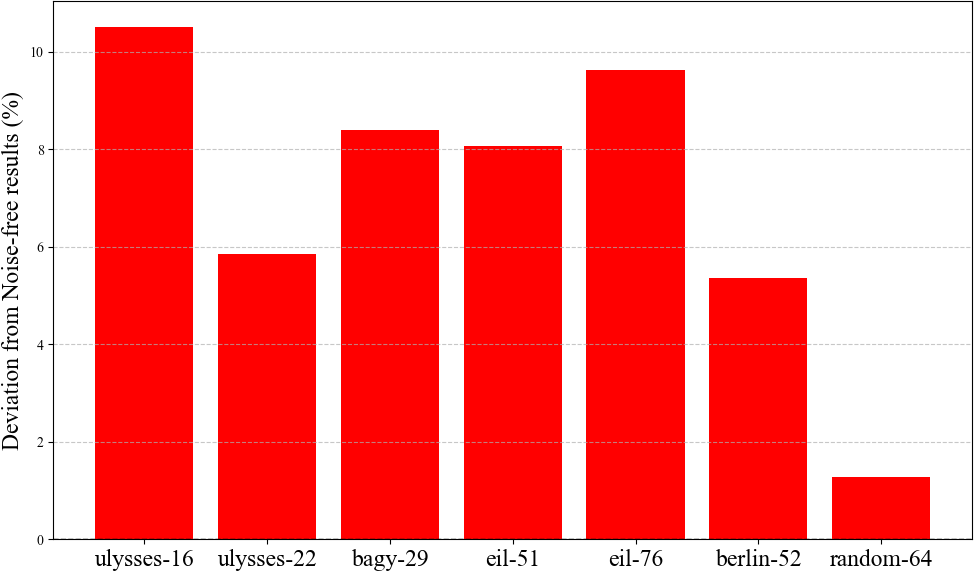}
}
\caption{The deviation of different noises: (a) Bit-flip Noise (b) Thermal Relaxation Noise}\label{noise}
\end{figure}

Since the current quantum computing hardware could not fully mitigate the effects of the noise, we also incorporate two types of noise
into the algorithm with Qiskit which is a quantum simulation tool developed by IBM.
The test results for a series of data sets with bit-flip and thermal relaxation noise effects are shown in table \ref{NoiseTab1} and
table \ref{NoiseTab2}.
The distances for the data sets, calculated with bit-flip noise and thermal relaxation noise at levels of 0.1\%, 1\%, 2\%, 5\%, and
10\%, are close to the ideal simulation results. The deviation comparisons are shown in figure \ref{noise}.
As seen from figure \ref{noise} and tables \ref{NoiseTab1} and \ref{NoiseTab2}, the deviations are not directly related to the
intensity of the noise, nor is it clearly related to the absolute magnitudes of the result values.
Similar behavior has also been observed when we incorporate random noise into the algorithm at different levels up to 10\%. These
results indicate that noise correction mechanism used in the extended
QACO algorithm effectively mitigates the impact of random noise on the computation results up to 10\%. This phenomenon is induced by
the Hamming distance method which corrects random noise induced invalid results.

\begin{table}[h]\centering
\caption{The results of data sets with bit-flip noises.}\label{NoiseTab1}
\begin{tabular}{ccccccc}
\hline
           & \multicolumn{6}{c}{Ratio of Noise} \\
Data-set   & 0.1 \%     & 1\%        & 2\%        & 5\%        & 10\%      &Deviation (\%) \\ \hline
ulysses-16 & 63.85      & 62.98      & 66.61      & 69.19      & 64.26     & 6.31          \\
ulysses-22 & 96.873     & 101.58     & 95.37      & 98.61      & 94.56     & 5.96          \\
bayg-29    & 14710.74   & 15137.13   & 13092.58   & 13941.60   & 12759.88  & 13.19         \\
eil-51     & 592.01     & 604.83     & 587.88     & 549.32     & 553.65    & 7.73          \\
eil-76     & 724.99     & 758.65     & 721.58     & 720.55     & 716.35    & 3.01          \\
berlin-52  & 10000.87   & 9199.02    & 9557.36    & 9749.5     & 9552.67   & 5.73          \\
random-64  & 149561.07  & 150896.13  & 150012.81  & 149881.99  & 151000.23 & 1.47          \\ \hline
\end{tabular}
\end{table}

\begin{table}[h]\centering
\caption{The results of data sets with thermal relaxation noises.}\label{NoiseTab2}
\begin{tabular}{ccccccc}
\hline
           & \multicolumn{6}{c}{Ratio of Noise} \\
Data-set   & 0.1 \%     & 1\%        & 2\%        & 5\%        & 10\%      &Deviation (\%)  \\ \hline
ulysses-16 & 63.91      & 70.08      & 69.15      & 71.47      & 65.63     & 10.52          \\
ulysses-22 & 100.11     & 98.43      & 93.28      & 99.54      & 94.12     & 5.85           \\
bayg-29    & 13291.67   & 14245.42   & 12446.39   & 13505.17   & 13630.71  & 8.40           \\
eil-51     & 558.97     & 587.95     & 600.02     & 600.57     & 560.37    & 8.07           \\
eil-76     & 748.72     & 799.67     & 798.76     & 800.55     & 722.96    & 9.64           \\
berlin-52  & 9200.21    & 10001.87   & 9637.20    & 9670.83    & 9159.82   & 5.36           \\
random-64  & 149783.23  & 151000.23  & 149121.31  & 149907.59  & 150001.03 & 1.28           \\ \hline
\end{tabular}
\end{table}

Median error rate for IBM quantum processor with single qubit gate error is approximately $0.03\%$ and
double-qubit controlled-Z gates is approximately $0.32\%$\cite{HeLH}.
If considering connectivity, the transpiled circuit will be much deeper and the error rate will be much higher. We assessed the error
rate based on Heron architecture, is about 0.26\%. This is still much lower than 0.32\%. This indicated our proposed algorithm can be
run in current NISQ devices effectively.

\begin{equation}
s = 1 - \prod_{j=1}^d \left( 1-\left(\frac{\sum_i E_{r_i} N_i}{\sum_i N_i}\right)_j\right)^{m_j}
\end{equation}
where, $N_i$  is the number of a quantum logic gate, $E_{r_i}$ is the corresponding error rate of this type of gate, d is the depth of
the quantum circuit, the last term, $\left(\frac{\sum_i E_{r_i} N_i}{\sum_i N_i}\right)$ is the average error rate in the j-th layer,
and $m_j$ is the number of gate at circuit layer j.
In our current study, we employed 5 nodes and 9 qubits to validate the effectiveness of the QACO algorithm with K-means clustering.
To further assess the
scalability of this approach, we can increase the number of nodes to 10, which would require 20 qubits. In this scenario, the
complexity of the quantum circuit and the error rate would increase accordingly. Based on the single-qubit and double-qubit gate error
rates of the IBM Heron quantum processor, we estimate that the noise error rate for a 10-node configuration would be approximately
between 0.5\% and 1.0\%.
If considering connectivity between the qubits, the noise level would be higher, but still no more than 10\% by rough estimation.
This suggests that even with an increased number of nodes, the extended QACO method maintains a relatively low noise error.
As we scale up the number of nodes and qubits, our method remains feasible for tackling problems involving hundreds of nodes.
This implies that the combined K-means clustering and QACO approach can potentially solve a wide range of complex combinatorial
optimization problems, such as production scheduling, path optimization, and more\cite{Jberclaz, LDeng}.

\section{Conclusions and future work}
This paper introduces a novel QACO algorithm enhanced with K-means clustering to overcome the limitations of current quantum computing hardware, particularly the scarcity of qubits and the presence of noise, in solving discrete optimization problems such as the TSP. The algorithm was tested using multiple benchmark datasets, and the results show that QACO with K-means clustering is not only an effective optimization method but also outperforms the traditional ACO algorithm. Furthermore, the proposed QACO algorithm demonstrates an ability to mitigate the impact of noise in existing quantum hardware, thereby significantly expanding the potential application scope of QACO.

 The number of subgroup nodes (corresponding to the number of qubits) and the number of groups determined by K-means can be dynamically adjusted based on the complexity of the problem at hand. This flexibility presents an opportunity for future optimization, allowing the algorithm to adapt to different problem scales and data structures. By increasing the number of qubits when dealing with larger datasets, we can potentially improve the computational efficiency and convergence speed of the algorithm. Meanwhile, adjusting the number of K-means clusters can be done to match the complexity and noise level of the data, ensuring optimal clustering performance and noise resilience. This dynamic adjustment of the number of qubits and K-means groups provides room for further improvements in the proposed algorithm. Future work could explore more sophisticated noise models or error correction mechanisms, potentially increasing the accuracy and robustness of the method. As quantum hardware continues to evolve, especially with advancements in qubit scaling and noise reduction, we expect the performance of the algorithm to improve in tandem. This lays out a promising direction for future research and optimizations as quantum computing technology matures.

\section{Acknowledgements}
This work is supported the Anhui Provincial Key Research and Development Project under Grant No. 2022b13020002, the Major Program of the Education Department of Anhui Province under Grant No. 2022AH040289, the Anhui Provincial Candidates for Academic and Technical Leaders Foundation under Grant No. 2019H208.
and YiWei Quantum Technology Co., Ltd.

\end{document}